\documentclass[aps,prl,twocolumn,superscriptaddress]{revtex4-1}

\usepackage[utf8]{inputenc}
\usepackage[english]{babel}
\usepackage[T1]{fontenc}
\usepackage{amsmath, amsfonts, amssymb, amsthm, mathrsfs, amssymb, braket, bbm, bm,graphicx,times,txfonts,color,subfigure}
\usepackage{hyperref}
\newcommand{\tr}{\mathrm{Tr}}
\newcommand{\CR}{{{\mathscr{C}}_{\mathcal{R}}}}
\newcommand{\CS}{{{\mathscr{C}}_{\mathcal{S}}}}
\newcommand{\CL}{{{\mathscr{C}}_{\ell_1}}}
\newcommand{\proj}[1]{\ket{#1}\!\bra{#1}}

\newcommand{\cU}{\mathcal{U}}

\begin{document}

\title{Robustness of coherence: An operational and observable measure of quantum coherence}

\author{Carmine~Napoli}
\affiliation{Dipartimento di Fisica ``E. R. Caianiello'', Universit\`a degli Studi di Salerno, Via Giovanni Paolo II, I-84084 Fisciano (SA), Italy; and INFN Sezione di Napoli, Gruppo Collegato di Salerno, Italy}
\affiliation{School of Mathematical Sciences, The University of Nottingham, University Park, Nottingham NG7 2RD, UK}

\author{Thomas~R.~Bromley}
\affiliation{School of Mathematical Sciences, The University of Nottingham, University Park, Nottingham NG7 2RD, UK}

\author{Marco~Cianciaruso}
\affiliation{Dipartimento di Fisica ``E. R. Caianiello'', Universit\`a degli Studi di Salerno, Via Giovanni Paolo II, I-84084 Fisciano (SA), Italy; and INFN Sezione di Napoli, Gruppo Collegato di Salerno, Italy}
\affiliation{School of Mathematical Sciences, The University of Nottingham, University Park, Nottingham NG7 2RD, UK}

\author{Marco~Piani}
\affiliation{SUPA and Department of Physics, University of Strathclyde, Glasgow G4 0NG, UK}

\author{Nathaniel~Johnston}
\affiliation{Department of Mathematics and Computer Science, Mount Allison University, Sackville, New Brunswick E4L 1E2, Canada}

\author{Gerardo~Adesso}
\affiliation{School of Mathematical Sciences, The University of Nottingham, University Park, Nottingham NG7 2RD, UK}

\begin{abstract}
 Quantifying coherence  is an essential endeavour for both quantum foundations and quantum technologies. Here the robustness of coherence is defined and proven a full  monotone in the context of the recently introduced resource theories of quantum coherence. The measure is shown to be observable, as it can be recast as the expectation value of a coherence witness operator for any quantum state. The robustness of coherence is evaluated analytically on relevant classes of states, and an efficient semidefinite program that computes it on general states is given.
An operational interpretation is finally provided: the robustness of coherence quantifies the advantage enabled by a quantum state in a phase discrimination task.
\end{abstract}


\date{March 1, 2016}
\maketitle


Nearly one century old, quantum mechanics is now livelier than ever. Fundamental experiments have just demonstrated, beyond any major loophole, that quantum correlations are incompatible with a local realistic interpretation \cite{loopholefree1,loopholefree2,loopholefree3}. Moreover, the realization that quantum properties can be harnessed for practical applications~\cite{nielsen2010quantum} is presently fuelling a heated international race~\cite{economist} to deploy quantum technologies \cite{QTech}. This is no coincidence: the improved comprehension of fundamental quantum properties and our increased ability to exploit them go hand in hand.

The most essential feature signifying quantumness in a single system and underpinning all forms of quantum correlations in composite systems~\cite{Horodecki2009,Modi2012,Brunner2014} is  {\em quantum coherence}, namely the possibility of creating superpositions of a set of orthogonal states. Revealing quantum coherence in the state of a natural or man-made system earmarks its behaviour as genuinely nonclassical \cite{Lloyd2011,Huelga2013}. Its degree of coherence may quantify the capability of such an object for quantum-enhanced applications \cite{Aaberg2006,Baumgratz2014}, ranging from cryptography \cite{Grosshans2003} to metrology \cite{Giovannetti2011} and thermodynamics  \cite{Lostaglio2015, Narasimhachar2015}.
It is thus imperative to accomplish a rigorous operational  characterization of quantum coherence.

Recently, various approaches have been put forward to develop a resource theory of  coherence \cite{Aaberg2006,Marvian2013,Baumgratz2014,Levi2014,Marvian2015,Streltsov2015b,Du2015,Hall2015,India2015,Winter2015,MaxCoh,Streltsov2015}. These partly follow from, and complement, earlier studies on resource theories of {\it asymmetry} \cite{Vaccaro2003,Bartlett2007,Vaccaro2008,Gour2008,Gour2009,Marvian2013,Marvian2014a}, of which coherence may be seen as a special instance \cite{Marvian2015,PRA}.
A resource theory is  defined by the notions of free states (i.e.~those not containing the resource, and assumed available at no cost) and free operations (i.e.~those one is restricted to, and that cannot transform free states into resource 
 states) \cite{Coecke2014,Brandao2015}. Fixing a reference basis (based on physical arguments \cite{Frozen}), which we can identify as the computational basis $\{\ket{j}\}_{j=0}^{d-1}$ for a $d$-dimensional system, the convex set ${{\mathscr{I}}}$ of free states in any resource theory of  coherence is given by incoherent states
 diagonal in the reference basis, $\delta = \sum_{j=0}^{d-1} \delta_j \ket{j}\!\bra{j}$.
 Any state $\rho$ can be reduced to an incoherent one by a full dephasing operation $\Delta$, which maps  $\rho$ into its diagonal part $\Delta(\rho) = \sum_{j=0}^{d-1} \ket{j}\!\bra{j} \rho \ket{j}\!\bra{j}$ in the reference basis.

Different authors have however considered different options in analyzing  limitations on the processing of coherence (see also \cite{Eric2016,MarvianSpekkens2016}). We mention the following alternative choices of free operations, in order of inclusion: incoherence preserving operations \cite{Aaberg2006} $\supset$ incoherent operations  \cite{Baumgratz2014} $\supset$ strictly incoherent operations \cite{YadinG2015} $\supset$ translationally invariant operations \cite{Marvian2015} $\supset$ genuinely incoherent operations \cite{Streltsov2015}. By incoherence preserving operations we refer to the maximal set of quantum channels $\Lambda^M$ which map incoherent states into incoherent states \cite{Aaberg2006}, i.e. $\Lambda^M (\delta)\in{{\mathscr{I}}}$ for any $\delta \in  {{\mathscr{I}}}$.
Incoherent operations  are instead those quantum channels $\Lambda^I$ which admit one operator-sum decomposition $\Lambda^I (\rho) = \sum_l K_l \rho {K_l}^\dagger$ with all incoherence preserving Kraus operators  $\{K_l\}$ \cite{Baumgratz2014}. Strictly incoherent operations $\Lambda^S$ are a subset of incoherent operations whose incoherence preserving Kraus operators  $\{K_l\}$ further obey $\langle j | K_l \rho K_l^\dagger |j\rangle  = \langle j | K_l \Delta(\rho) K_l^\dagger |j\rangle$ $\forall j,l$, meaning that they can neither create nor use coherence \cite{YadinG2015,Levi2014}. More restrictively, genuinely incoherent operations $\Lambda^G$ \cite{Streltsov2015} (also known as generalized incoherent measurements \cite{YadinG2015}) are those which leave  every incoherent state invariant,  $\Lambda^G(\delta) = \delta$ \cite{Streltsov2015}; their Kraus operators are all incoherence preserving in all possible operator-sum decompositions. In-between the last two sets are translationally invariant operations, introduced within the resource theory of asymmetry \cite{Marvian2013,Marvian2015}: specialized to coherence (i.e., asymmetry with respect to time translations generated by a Hamiltonian $H$ diagonal in the reference basis $\{\ket{j}\}$), translationally invariant operations $\Lambda^T$ are defined by the  condition
$\text{e}^{-{\rm i} H t} \Lambda^T(\rho) \text{e}^{{\rm i} H t} = \Lambda^T(\text{e}^{-{\rm i} H t} \rho \text{e}^{{\rm i} Ht}) $ for any  $\rho$ and any real $t$.  

Several quantities have been proposed accordingly as candidate measures of quantum coherence, respecting physical requirements of monotonicity under (some of) the sets of operations introduced above \cite{Aaberg2006,Vaccaro2008,Marvian2013,Baumgratz2014,Levi2014,Marvian2015,Streltsov2015b,Du2015,Winter2015,Streltsov2015,MaxCoh,Marvian2014a,Girolami2014,Frozen}. A canonical measure which complies with all the aforementioned resource theories is the relative entropy of coherence \cite{Aaberg2006,Vaccaro2008,Marvian2013,Baumgratz2014}, which for a state $\rho$ takes the simple form $\CS(\rho) = {\cal S}(\Delta(\rho))-{\cal S}(\rho)$, where ${\cal S}(\rho) = -\tr[\rho \log_2 \rho]$ is the von Neumann entropy.  This measure can be interpreted as the optimal rate of maximally coherent states that can be distilled by incoherent operations $\Lambda^I$ in the asymptotic limit of many copies of $\rho$ \cite{Winter2015}; however its experimental determination requires full state tomography, which can be unfeasible for large systems. More accessible measures of relevance for quantum metrology \cite{Giovannetti2011}, such as the Wigner-Yanase skew information and the quantum Fisher information \cite{Girolami2014,Marvian2014a}, are monotone under translationally invariant operations  but not under the larger set of incoherent operations  \cite{Marvian2015}, which may put into question their broader status as coherence quantifiers. In general, despite recent progress, there remains a shortage of  rigorous and physically intuitive bona fide measures of coherence endowed with direct operational interpretations.

In this Letter we fill this gap by introducing the {\em robustness of coherence}. As the name suggests, it quantifies the minimal mixing required to destroy all the coherence in a quantum state --- an already operational definition, inspired by similar concepts previously investigated for entanglement, steering-type correlations, non-locality and other resources~\cite{Robustness,GenRobustness,Piani2015,GellerPiani,Brandao2015}. We prove that such a measure is a full monotone in all possible resource theories of coherence. The measure is furthermore computable (exactly in relevant cases, and numerically in general via a simple semidefinite program  \cite{epapsL}) and observable: it can be recast as the expectation value of a witness operator for any quantum state. This makes it very appealing for experimental investigations, e.g.~in condensed matter and biological contexts \cite{Lloyd2011,Li2012,Huelga2013}. We then show that the measure admits a direct operational interpretation: it quantifies the advantage enabled by a quantum state, compared to any incoherent state, in a phase discrimination task. We further discuss the generalization of these results to the case of asymmetry in a companion paper \cite{PRA}, which also contains detailed proofs for some technical results of this Letter.

\begin{figure}[t!]
  \centering
{\includegraphics[width=4.5cm]{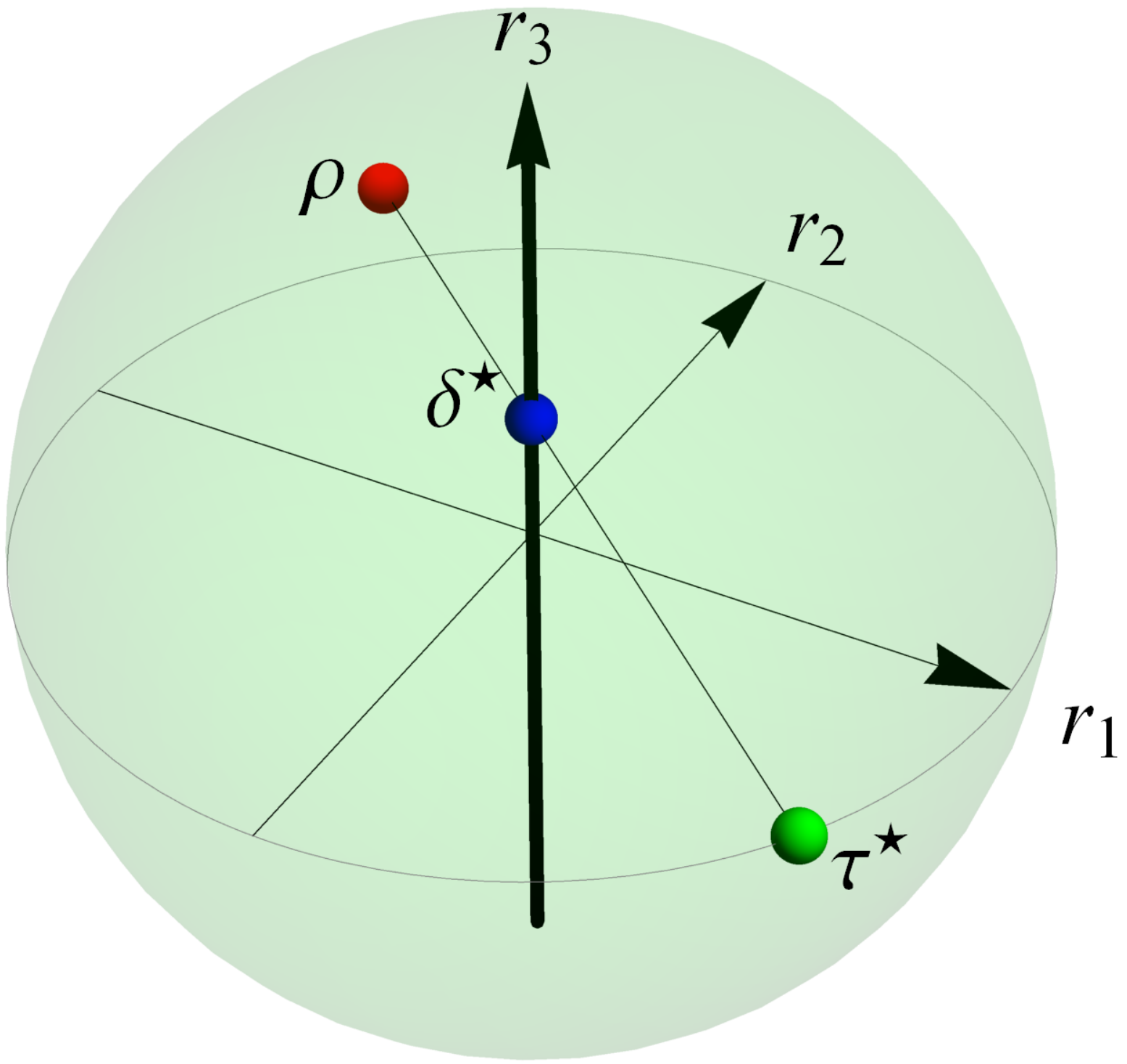}}
 \caption{ \label{FigQ}
(Color online) Robustness of coherence $\CR(\rho)$ for a single qubit state $\rho = \frac12 (\openone + \vec{r} \cdot \vec{\sigma})$, where $\vec{r}$ is the Bloch vector and $\vec\sigma$ is the vector of Pauli matrices. Incoherent states span the thick vertical $r_3$ axis. The optimization in Eq.~(\ref{R}) is fulfilled by an equatorial pure state $\tau^{\star}$ as depicted, resulting in $\CR(\rho) =  (r_1^2+r_2^2)^{\frac12} = 2|\rho_{01}|$.
}
\end{figure}

Let ${\mathscr{D}}(\mathbb{C}^d)$ be the convex set of density operators acting on a $d$-dimensional Hilbert space, and let $\mathscr{I} \subset {\mathscr{D}}(\mathbb{C}^d)$ be the subset of incoherent states. We define the robustness of coherence (RoC) of a quantum state $\rho \in {\mathscr{D}}(\mathbb{C}^d)$ as
\begin{equation}\label{R}
\CR(\rho) = \min_{\tau \in {\mathscr{D}}(\mathbb{C}^d)} \left\{ s\geq 0\ \Big\vert\ \frac{\rho + s\ \tau}{1+s} =: \delta \in {\mathscr{I}}\right\}\,,
\end{equation}
that is, the minimum weight of another state $\tau$ such that its convex mixture with $\rho$ yields an incoherent state $\delta$.
The concept is illustrated in Fig.~\ref{FigQ} for a qubit ($d=2$). If we denote by $\tau^\star$ and $\delta^\star$ the states achieving the minimum in (\ref{R}), then
\begin{equation}\label{pseudo}
\rho = \big(1+\CR(\rho)\big)\delta^\star - \CR(\rho)\tau^\star\,,
\end{equation}
is said to realize an optimal pseudomixture for $\rho$.
Notice that it is necessary in Eq.~(\ref{R}) to let $\tau$ be an arbitrary state: if one restricted $\tau$ to be incoherent, then the minimum $s$ would diverge for any state $\rho$ with nonzero coherence, henceforth resulting totally uninformative. This contrasts with the case of entanglement, for which the original robustness was defined in terms of pseudomixtures with separable states \cite{Robustness}, and only later extended to pseudomixtures with arbitrary states \cite{GenRobustness}.

We now prove that the RoC is a bona fide measure of coherence. First of all, it is seen by definition that \begin{equation}\label{C1}
\mbox{$\CR(\rho) \geq 0$\quad and \quad $\CR(\rho) = 0\ \ \Longleftrightarrow\ \ \rho \in \mathscr{I}$}\,.
  \end{equation}

  Second, the RoC is convex, which is a desirable (although not fundamental) property for a coherence quantifier \cite{Baumgratz2014}. The proof mirrors the one for the robustness of entanglement \cite{Robustness}. Let $\rho_1$ and $\rho_2$ be two states, and write for each the optimal pseudomixture  $\rho_k = \big(1+\CR(\rho_k)\big)\delta_k^\star - \CR(\rho_k) \tau_k^\star$ ($k=1,2$). Taking the convex combination $\rho = p \rho_1 + (1-p) \rho_2$ with $0 \leq p \leq 1$, notice that a  pseudomixture  $\rho = (1+s) \delta - s \tau$ can be written, with $\delta =
\big[p\big(1+\CR(\rho_1)\big)\delta^\star_1 + (1-p)\big(1+\CR(\rho_2)\big)\delta^\star_2\big]/(1+s) \in \mathscr{I}$, $\tau = \big[p \CR(\rho_1) \tau^\star_1 + (1-p) \CR(\rho_2) \tau^\star_2\big]/s$, and $s = p \CR(\rho_1) + (1-p) \CR(\rho_2)$. By definition,  $\CR(\rho) \leq s$, which proves convexity,
\begin{equation}
\label{C3}
\CR\big(p \rho_1 + (1-p) \rho_2\big) \leq  p \CR(\rho_1) + (1-p) \CR(\rho_2)\,.
\end{equation}

Third, and most importantly, the RoC is nonincreasing under all the sets of operations used in resource theories of coherence. We prove in fact a general form of monotonicity under incoherence preserving (sub)channels. Let $\{\Gamma_l\}_{l=1}^m$ be an instrument, i.e., a set of $m$ (sub)channels (completely positive maps whose sum $\sum_{l=1}^m \Gamma_l(\rho) =: \Lambda(\rho)$ defines a trace preserving channel $\Lambda$), mapping any incoherent state $\delta \in \mathscr{I}$ into another (un)normalized incoherent state $\Gamma_l(\delta)$. For any  $\rho$, we have then
\begin{equation}\label{C2}
\CR(\rho) \geq \sum_{l=1}^m \tr[\Gamma_l(\rho)]\ \CR \left(\frac{\Gamma_l(\rho)}{\tr[\Gamma_l(\rho)]} \right)\,.
\end{equation}
The proof can be easily sketched (see \cite{PRA} for more details).  Take the optimal pseudomixture for $\rho$ given by Eq.~(\ref{pseudo}) and apply the (sub)channel $\Gamma_l$ to both sides, $\Gamma_l(\rho) = \big(1+\CR(\rho)\big)\Gamma_l(\delta^\star) - \CR(\rho)\Gamma_l(\tau^\star)$.
Since $\Gamma_l(\delta^\star)/{\tr[\Gamma_l(\delta^\star)]}$ is still incoherent,  definition (\ref{R}) implies $\CR\left(\frac{\Gamma_l(\rho)}{\tr[\Gamma_l(\rho)]}\right) \leq \CR(\rho)\frac{\tr[\Gamma_l(\tau^\star)]}{\tr[\Gamma_l(\rho)]}$. Then, $\sum_l {\tr[\Gamma_l(\rho)]} \CR\left(\frac{\Gamma_l(\rho)}{\tr[\Gamma_l(\rho)]}\right) \leq \sum_l {\tr[\Gamma_l(\rho)]} \CR(\rho)\frac{\tr[\Gamma_l(\tau^\star)]}{\tr[\Gamma_l(\rho)]}
 = \CR(\rho) \sum_l {\tr[\Gamma_l(\tau^\star)]} = \CR(\rho)$, concluding the proof.
In the case $m=1$,  Eq.~(\ref{C2}) proves that the RoC cannot increase, on average, under the maximal set of incoherence preserving operations $\{\Lambda^M\}$ \cite{Aaberg2006}. For $m \geq 1$, if one identifies each $\Gamma_l$ with a Kraus operator $K_l$ (obeying $\sum_{l=1}^{m} {K_l}^{\dagger} K_l = \openone$), then Eq.~(\ref{C2}) proves monotonicity of the RoC under selective incoherent operations $\{\Lambda^I\}$, flagged as property C2b in \cite{Baumgratz2014}, which is typically a rather demanding requirement in resource theories. Overall,  Eq.~(\ref{C2}) establishes the RoC as a {\it full monotone} with respect to {\it all} possible formulations of the theory of coherence.


We now show that the RoC has also desirable properties of computability and  accessibility.
Inspired by entanglement witnesses \cite{Horodecki2009,QWitness1}, which are very useful tools to detect inseparability in laboratory \cite{GuhneToth}, we introduce the notion of \emph{coherence witnesses}. A Hermitian operator $W$ satisfies $\Delta(W) \geq 0$ if and only if $\tr[\delta W] = \tr[\delta \Delta(W)] \geq 0$ for all incoherent states $\delta \in \mathscr{I}$; we call any such observable $W$ a coherence witness, because finding $\tr[\rho W]<0$ reveals coherence in the state $\rho$ \cite{Note1}.
We find that the expectation value of any witness $W$, obeying the further constraint $W \leq  \openone$,  provides a quantitative lower bound to the RoC \cite{PRA},
\begin{eqnarray}\label{W}
\max\{0,\,-\tr[\rho W]\} \leq \CR(\rho)\,,\ \ && \forall\ W \mbox{ such that} \\
\label{constraints}
&&\Delta(W) \geq 0  \mbox{ and }   W \leq  \openone\,.
\end{eqnarray}

Importantly, given any state $\rho$, there always exists an optimal witness $W^\star$, characterized in particular by $\Delta(W^\star)=0$, which saturates inequality (\ref{W}). In other words, the RoC is an {\it observable} quantity, given by the expectation value of a suitable (state-dependent) witness operator for any quantum state $\rho$. Finding such an optimal witness, hence determining $\CR(\rho)$ as defined in (\ref{R}), can be then recast \cite{PRA} as a simple semidefinite program~\cite{vandenberghe1996semidefinite} (significantly more efficient than the convex optimization one for the robustness of entanglement \cite{Brandao2005}):
\begin{equation}\label{semi}
\mbox{maximize \quad $-\tr[W \rho]$ \quad subject to \quad  Eq.~(\ref{constraints})}\,.
\end{equation}
For the convenience of the reader, we release MATLAB~\cite{MATLAB:2015} code that makes use of the free CVX package~\cite{cvx,gb08} to evaluate the RoC, as a Supplemental Material \cite{epapsL}.

These results reveal that one can readily estimate the RoC from below in laboratory, by measuring any observable $W$ obeying the constraints in (\ref{constraints}), with no need for full tomography of the state $\rho$. This may be particularly valuable for witnessing coherence effects in biological domains, e.g.~energy transport phenomena in light-harvesting systems \cite{Lloyd2011,Li2012,Huelga2013,Whaley2010}.
 However, given a state $\rho$, the lower bound of Eq.~(\ref{W}) can vanish for non-optimized choices of $W$. Typically, one needs some knowledge on the form of $\rho$ to determine the optimal witness $W^\star$; a similar issue is encountered in entanglement detection  \cite{GuhneToth}. Nonetheless Eqs.~\eqref{W} and \eqref{constraints} imply that, for any set of observables $\{O_i\}$, $i=1,\ldots,k$, experimentally measured with expectation values $o_i=\tr[O_i\rho]$, and not necessarily tailored to the measurement of RoC,
one can consider coherence witnesses of the form $W=\sum_{i=1}^k c_i O_i + m\openone$, for $c_1,\ldots,c_k,m \in\mathbb{R}$, and obtain a lower bound to the RoC  by the SDP~\cite{PRA}
(code available~\cite{epapsL})
\begin{align*}
\textrm{maximize}\quad&\mbox{$-\big(\sum_{i=1}^k c_i o_i + m\big)$}\\
\textrm{subject to}\quad&\mbox{$\Delta\big(\sum_{i=1}^k c_i O_i + m\openone\big)\geq 0,\quad \sum_{i=1}^k c_i O_i + m\openone \leq \openone.$}
\end{align*}
One can even make potentially better use of available experimental data, by exactly estimating the minimal RoC compatible with the data; this can also be cast as an SDP~\cite{PRA,epapsL}.

Accessible \emph{faithful} lower bounds to the RoC can be given too, noting that $W_2=(\Delta(\rho)-\rho)/\|\Delta(\rho)\|_\infty$ obeys \eqref{constraints}, so that
\begin{equation}
\label{eq:lowerbounds}
\CR(\rho) \geq \frac{\|\rho-\Delta(\rho)\|_2^2}{\|\Delta(\rho)\|_\infty}\geq\frac{\|\rho-\Delta(\rho)\|_2^2}{\|\Delta(\rho)\|_2}\geq \|\rho-\Delta(\rho)]\|_2^2,
\end{equation}
since $\tr[(\Delta(\rho)-\rho)\rho]=\tr[\Delta(\rho)^2]-\tr[\rho^2]=\|\rho-\Delta(\rho)\|_2^2$. Here, $\|\cdot\|_2$ is the 2-norm, and $\|\cdot\|_\infty$ is the operator norm. The quantity on the rightmost-hand side of \eqref{eq:lowerbounds} is: (i) nonzero on all but incoherent states; (ii) itself a monotone under genuinely incoherent operations $\Lambda^G$ \cite{Streltsov2015}, but not under the larger sets of incoherent operations  \cite{Baumgratz2014}; (iii) accessible via the measurement of the purities $\tr[\rho^2]$ and $\tr[\Delta(\rho)^2]$ (notably, the same holds for the tighter second-to-last bound in \eqref{eq:lowerbounds}). The latter two quantities can be measured directly on two copies of the state $\rho$ (assumed unknown), as
$\tr[\rho^{\otimes 2} V]$ and $\tr[\rho^{\otimes 2}\Delta^{\otimes 2}(V)]$, respectively, with $V$ being the swap operator~\cite{ekert2002direct,Girolami2014}, defined by its action $V \ket{\psi} \ket{\phi} = \ket{\phi} \ket{\psi}$, for all $\ket{\psi}, \ket{\phi} \in \mathbb{C}^{d}$.

We now show that an {\it analytical} evaluation of RoC can be obtained for a relevant class of $d$-dimensional states. Let $\rho \in {\mathscr{D}}(\mathbb{C}^d)$ be a state for which there exists a unitary $U = \sum_j \text{e}^{{\rm i} \phi_j} \ket{j}\!\bra{j}$, belonging to the set of genuinely incoherent operations \cite{Streltsov2015}, such that $(U \rho U^\dagger)_{kl} = |\rho_{kl}|$. One has then
$\CR(\rho) = \CL(\rho)$ \cite{PRA},
where 
$
\CL(\rho) = \sum_{k,l} |\rho_{kl}|-1 = 2 \sum_{k<l} |\rho_{kl}|$ is the $\ell_1$ norm of coherence \cite{Baumgratz2014}. 
The class of states for which this equality 
holds includes, for instance, all one-qubit states ($d=2$, see Fig.~\ref{FigQ}), all $d$-dimensional states with X-shaped density matrix \cite{SaiX,Hashemi2012,noteX} (which contain in particular Bell diagonal states of two qubits  \cite{Tstates,Frozen}),
 and all pure states $\ket{\psi} \in \mathbb{C}^d$. For the latter, writing in general $\ket{\psi} = \sum_{j=0}^{d-1} \psi_j \ket{j}$, we get explicitly $\CR(\proj{\psi}) = \CL(\proj{\psi}) = \big(\sum_j |\psi_j|\big)^2-1$ \cite{Baumgratz2014}.

In particular, maximally coherent states $\ket{\psi^+}$, characterized by $|\psi_j| = \frac{1}{\sqrt{d}}\ \forall j=0,\ldots,d-1$, have $\CR(\proj{\psi^+}) = \CL(\proj{\psi^+}) = d-1$, that is the maximum possible value for the RoC of any $d$-dimensional state.  One can show~\cite{PRA} in fact that these are the only states which can reach maximal RoC, which positively settles another requirement recently advocated for bona fide measures of coherence \cite{MaxCoh}.

The equivalence between RoC and $\ell_1$ norm of coherence breaks down already in dimension $d=3$. One can prove however the existence of general upper and lower bounds~\cite{PRA},
\begin{equation}\label{bounds}
(d-1)^{-1}\CL(\rho) \leq \CR(\rho) \leq \CL(\rho)\,,\quad \forall \ \rho \in {\mathscr{D}}(\mathbb{C}^d)\,.
\end{equation}
Both bounds can be  tight. Examples of states saturating the upper bound
have been provided already (for instance, all pure states). A family of states saturating the lower bound is instead given by
$\rho_p=(1+p)\openone/d-p\proj{\psi^+}$, with $0\leq p \leq \frac{1}{d-1}$, for which $\CL(\rho_p)=p(d-1)$ and $\CR(\rho_p)= p$.
Nonetheless, the lower bound becomes looser for large values of $\CL$, and one finds $\CR(\rho)\rightarrow d-1$ for all $\rho$ such that $\CL(\rho) \rightarrow d-1$~\cite{PRA}.

We are finally ready to provide a direct operational interpretation for the RoC in a metrology context. Consider the following {\it phase discrimination} (PD) game. Alice prepares a quantum state $\rho \in {\mathscr{D}}(\mathbb{C}^d)$, which then enters a black box. The black box encodes a phase on $\rho$ by implementing a unitary $U_\phi=\exp({\rm i}N\phi)$, with $N=\sum_{j=0}^{d-1} j \proj{j}$ and $\phi\in\mathbb{R}$, so that the output state is determined by the action of the unitary channel $\mathcal{U}_\phi(\rho):=U_\phi\rho U^\dagger_\phi$. We can think of $N$ as a  Hamiltonian for the system with equispaced spectrum, assuming unit spacing without loss of generality. In this way, the reference basis $\{\ket{j}\}$, with respect to which coherence is defined and measured, is physically identified by the choice of the Hamiltonian. Suppose one of $m$ phases $\{\phi_k\}_{k=0}^{m-1}$ can be applied, each with a prior probability $p_k$. Any collection of pairs $\{(p_k,\phi_k)\}_{k=0}^{m-1}=:\Theta$ defines a PD game, where Alice's goal is that of guessing correctly the phase that was actually imprinted on the state. To this end, she performs a generalized measurement with elements $\{M_k\}$ (satisfying $M_k \geq 0$, $\sum_k M_k = \openone$) on the output state $\mathcal{U}_\phi(\rho)$ after the black box. Optimizing over all measurements, the maximal probability of success 
depends on the game $\Theta$ and the input state $\rho$, and is given by 
$p^{\rm succ}_\Theta(\rho) = \max_{\{M_k\}} {\sum}_k p_k \tr[U_{\phi_k}\rho U^\dagger_{\phi_k} M_k]$ .

Suppose now Alice's input state is incoherent, $\rho \equiv \delta \in \mathscr{I}$. Since every unitary channel $\mathcal{U}_\phi$ leaves any such state invariant, $\cU_\phi(\delta)=\delta$, the best strategy for Alice is always to cast the guess $k^{\max}$ corresponding to the phase
with the highest prior probability $p_{k^{\max}}:=\max_k p_k$.
This results in an optimal probability of success for any incoherent state given by $p^{{\rm succ}}_{\Theta}(\mathscr{I}):= p_{k^{\max}}$, which can be achieved even without actually probing the channel, just by a fixed guess. 

It is clear that, by preparing a coherent state $\rho \notin \mathscr{I}$, Alice can expect to do better. What is less obvious yet more remarkable, is that the maximum advantage achievable by using $\rho$ as opposed to any incoherent probe $\delta$, in all possible PD games, is quantified exactly by the RoC of $\rho$. More precisely~\cite{PRA}:
\begin{equation}\label{teo}
\max_{\Theta} \frac{p^{\rm succ}_{\Theta}(\rho)}{p^{{\rm succ}}_{\Theta}(\mathscr{I})}  = 1 + \CR(\rho)\,.
\end{equation}
The maximum is achieved for the PD game $\Theta^\star\equiv\big\{\big(\frac1d,\frac{2\pi k}{d}\big)\big\}_{k=0}^{d-1}$.
Therefore $\CR(\rho)$ exactly quantifies, in particular, how useful the state $\rho$ is for reliable decoding and transmission of messages encoded by generalized phase channels $\rho\mapsto Z^k \rho Z^{\dagger k}$, with 
$Z \ket{j} = \exp\big({{\rm i}\frac{2\pi}{d}j}\big)\ket{j}$. These channels feature  in  several quantum information tasks such as quantum error correction \cite{Gottes99}, cloning \cite{Cerf2000}, and dense coding \cite{Hiroshima2001,Werner2001}.
This suggests a prominent role of coherence, specifically measured by the RoC, in quantum communication.

We notice  that one can consider more general channel discrimination (CD) games, where each game is associated with a set of pairs $\{(p_k,\Lambda_k)\}_{k=0}^{m-1}=:\Upsilon$, with $\{\Lambda_k\}$ a set of $m$ (generally nonunitary) channels. For each CD game $\Upsilon$, Alice's goal is still that of discriminating which $\Lambda_k$ gets  applied by a black box to an input $\rho$, and she succeeds with optimal probability $p^{\rm succ}_\Upsilon(\rho) = \max_{\{M_k\}} {\sum}_k p_k \tr[\Lambda_k(\rho) M_k]$, where we optimize over measurements similarly as before.  By virtue of Eq.~\eqref{pseudo}, for any CD game $\Upsilon$ it holds $p^{\rm succ}_\Upsilon(\rho) \leq (1+\CR(\rho))p^{\rm succ}_\Upsilon(\mathscr{I})$, where $p^{\rm succ}_\Upsilon(\mathscr{I})$ is the best probability of success by using as input any incoherent state. In general, $p^{\rm succ}_\Upsilon(\mathscr{I})$ can be higher than $p_{k^{\max}}$, because the channels $\Lambda_k$ may act nontrivially on incoherent states. Nonetheless, if one focuses on a subclass of CD games $\{\Upsilon^\star\}\ni \Theta^\star$ containing the PD game $\Theta^\star$, one gets $\max_{\Upsilon\in\{\Upsilon^\star\}} \frac{p^{\rm succ}_{\Upsilon}(\rho)}{p^{{\rm succ}}_{\Upsilon}(\mathscr{I})}  = 1 + \CR(\rho)$. The RoC $\CR(\rho)$ thus quantifies the maximum achievable advantage in any CD task in which the phase channels $Z^k$ are some of the possible channels applied to a probe $\rho$.
It will be a worthy development to extend this analysis to the scenario of {\it assisted} CD games, where the collaboration of a correlated party Bob may further increase Alice's probability of success in the discrimination \cite{Chitambar2015}.

We conclude by remarking that the definition (\ref{R}) can be extended to a more abstract notion of {\it robustness of asymmetry} \cite{PRA}, in which the free states (symmetric states) are those invariant under the action of a group  \cite{Marvian2013}. Specifically, given a symmetry group ${\sf G}$ with associated unitary representation  $\{U_g\}_{g \in {\sf G}}$ on the Hilbert space $\mathscr{H}$, and defining the action of $U_g$ on a state $\rho \in {\mathscr{D}}(\mathscr{H})$  as ${\cal U}_g(\rho) = U_g \rho U_g^\dagger$, a state $\sigma \in {\mathscr{D}}(\mathscr{H})$ is  symmetric with respect to $\sf G$ if and only if
${\cal U}_g(\sigma) =\sigma$ for all $g\in {\sf G}$. Denoting by $\mathscr{S}$ the convex set of all symmetric states, the robustness of asymmetry of a state $\rho$ is then defined as ${{{\mathscr{A}}_{\mathcal{R}}}}(\rho) = \min_{\tau \in \mathscr{D}(\mathscr{H})} \left\{ s\geq 0\ \big\vert\ \frac{\rho + s\ \tau}{1+s} =: \sigma \in {\mathscr{S}}\right\}$, i.e.,~as the minimum weight of another state $\tau$ such that its convex mixture with $\rho$ yields a symmetric state $\sigma$.
Then, suitable adaptations of all the properties demonstrated above in Eqs.~(\ref{C1})--(\ref{eq:lowerbounds}) carry over to the robustness of asymmetry, including the SDP evaluation and an operational interpretation in the context of channel discrimination games \cite{PRA}. Coherence can be recovered as a special case of  asymmetry with respect to the $d$-dimensional representation of the compact group ${\sf U}(1)$.

The approach pursued in this Letter, based on the generalized notion of robustness, appears accordingly quite versatile to define and validate insightful quantifiers of resources in quantum physics \cite{Brandao2015} and beyond \cite{Coecke2014,Renner2015}, as demonstrated here for the fundamental case of quantum coherence.

{\it Acknowledgments.} This work has received funding from the European Research Council (ERC StG GQCOP, Grant No.~637352),
from the European Union's Horizon 2020 research and innovation programme under the Marie Sklodowska-Curie (Grant Agreement No.~661338),
and the Erasmus+ Programme. We thank D.~Girolami, M.~Hall, I.~Marvian, M.~B.~Plenio, A.~Streltsov, and J.~Vaccaro for useful discussions.

\bibliographystyle{apsrevfixed}
\bibliography{corrub}

\begin{thebibliography}{66}
\bibitem [{not()}]{noteX}%
  \BibitemOpen
  \href@noop {} {}\bibinfo {note} {Generalized X-states can be defined as $\rho
  = \sum_{k=0}^{d/2}\rho_{k}$ if $d$ is even, and $\rho = \sum_{k=0}^{\lfloor
  d/2\rfloor}\rho_{k}+\rho_{c}$ if $d$ is odd, with $\rho_k =
  \rho_{kk}\proj{k}+\rho_{k,d-1-k}\ket{k}\bra{d-1-k}
  +\rho_{d-1-k,k}\ket{d-1-k}\bra{k}+\rho_{d-1-k,d-1-k}\proj{d-1-k}$, and
  $\rho_{c}= \rho_{\lfloor d/2\rfloor+1}\proj{\lfloor
  d/2\rfloor+1}$.}\BibitemShut {Stop}%
\end{thebibliography}

\end{document}